\documentclass[letter]{aa}
\usepackage{graphicx}
\usepackage{txfonts}
\usepackage{natbib}
\usepackage{url}
\usepackage{multirow,bigdelim,multicol}
\usepackage{rotating}
\usepackage{longtable}
\usepackage{ulem}
\usepackage{placeins}
\usepackage{color}
\usepackage{bm}

\begin{document}

\newcommand{\zabs}{\ensuremath{z_{\rm abs}}}
\newcommand{\zem}{\ensuremath{z_{\rm em}}}
\newcommand{\zqso}{\ensuremath{z_{\rm QSO}}}
\newcommand{\HI}{\ion{H}{i}}
\newcommand{\CI}{\ion{C}{i}}
\newcommand{\ClI}{\ion{Cl}{i}}
\newcommand{\ClII}{\ion{Cl}{ii}}
\newcommand{\ClH}{[\ClI/H$_2$]}
\def\h2{$\rm H_2$}
\newcommand{\cmsq}{\ensuremath{{\rm cm}^{-2}}}
\newcommand{\kms}{\ensuremath{{\rm km\,s^{-1}}}}
\newcommand{\qso}{Q$\,$1232+082}
\newcommand{\PN}{\color{red} PN:~}
\newcommand{\avg}[1]{\left< #1 \right>} 
\newcommand{\co}[2]{\sout{#1} {\color{blue} #2}}
\newcommand{\re}[2]{\sout{#1} {\color{red} #2}}
\newcommand{\SB}{\color{blue} SB:~}

\newcommand{\ioffe}{Ioffe Physical-Technical Institute of RAS, {Polyteknicheskaya 26}, 194021 Saint-Petersburg, Russia\label{ioffe}}
\newcommand{\sptg}{St.-Petersburg Polytechnic University, {Polyteknicheskaya 29}, 195251 Saint-Petersburg, Russia\label{sptg}}
\newcommand{\iap}{Institut d'Astrophysique de Paris, CNRS-UPMC, UMR7095, 98bis bd Arago, 75014 Paris, France\label{iap}}
\newcommand{\eso}{European Southern Observatory, Alonso de C\'ordova 3107, Vitacura, Casilla 19001, Santiago 19, Chile\label{eso}}
\newcommand{\iucaa}{Inter-University Centre for Astronomy and Astrophysics, Post Bag 4, Ganeshkhind, 411\,007 Pune, India\label{iucaa}}

\title{Neutral chlorine and molecular hydrogen at high redshift}
\titlerunning{Cl\,{\sc i} and H$_2$ at high-$z$}

\author{
     S.A.~Balashev      \inst{\ref{ioffe},\ref{sptg}}
\and P.~Noterdaeme       \inst{\ref{iap}}
\and V.V.~Klimenko       \inst{\ref{ioffe},\ref{sptg}}
\and P.~Petitjean        \inst{\ref{iap}}
\and \\
R.~Srianand         \inst{\ref{iucaa}}
\and C.~Ledoux           \inst{\ref{eso}}
\and A.V.~Ivanchik       \inst{\ref{ioffe},\ref{sptg}}
\and D.A.~Varshalovich   \inst{\ref{ioffe},\ref{sptg}}
       }

  \institute{    
\ioffe
\and \sptg
\and \iap
\and \eso
\and \iucaa
             }
\date{Accepted 28/01/2015, Received 19/12/2014}


\abstract{
Chlorine and molecular hydrogen are known to be tightly linked together in the cold 
phase of the local interstellar medium through rapid chemical reactions. 
We present here the first systematic study of this relation at high redshifts using H$_2$-bearing 
damped Ly$\alpha$ systems (DLAs) detected along quasar lines of sight.
Using high-resolution spectroscopic data from VLT/UVES and Keck/HIRES, we report
the detection of Cl\,{\sc i} in 9 DLAs (including 5 new detections) out of 18 high-$z$ DLAs 
with $N($H$_2) \ge 10^{17.3}$~cm$^{-2}$ (including 
a new H$_2$ detection at $z=3.09145$ towards J\,2100$-$0641) and present 
upper limits for the remaining 9 systems. 
We find a $\sim$5\,$\sigma$ correlation between 
$N$(Cl\,{\sc i}) and $N$(H$_2$) with only $\sim$0.2~dex dispersion over the range 
18.1~$<$~log~$N$(H$_2$)~$<$~20.1, thus probing column densities 10 times lower those seen
towards nearby stars, roughly following the relation 
$N$(Cl\,{\sc i}$) \approx 1.5\times10^{-6} \times N($H$_2)$. This relation between
column densities is surprisingly the same at low and high redshift
suggesting that the physical and chemical conditions are similar for a given H$_2$ 
(or \ClI) column density.
In turn, the $N(\mbox{\ClI})/N({\rm H_2})$ ratio is found to be uncorrelated with the {\sl overall} metallicity
in the DLA. 
Our results confirm that neutral chlorine is an excellent tracer of molecule-rich gas 
and show that the molecular fraction or/and metallicity in the H$_2$-bearing component of DLA 
could possibly be much higher than the line-of-sight average values usually measured in DLAs. 
}

\maketitle

\section{Introduction \label{introduction}}

It has been shown that the global star-formation rate in the Universe gradually increases
from z$\sim$10 to z$\sim$2-3 and then steeply decreases till the present epoch, 
$z=0$ \citep[see e.g.][and references therein]{Dunlop2011}. Because metals are 
produced by stars, the determination of metal abundance in the gas provides complementary information 
about star-formation history \citep{Rafelski2012}. This can be done using Damped Lyman-$\alpha$ 
absorption systems (DLAs) that represent the main reservoir of 
neutral gas at high redshift \citep{Prochaska2009, Noterdaeme2009} and are likely to be located 
in galaxies or in their close environment \citep[e.g.][]{Krogager2012}.
DLAs arise mostly in 
the warm neutral medium \citep[e.g.][]{Petitjean2000, Kanekar2014} and have a multicomponent velocity 
structure, with metal absorption lines spread typically over 100-500 km\,s$^{-1}$ 
\citep{Ledoux1998}. In a small fraction of DLAs, the line of sight intercepts 
cold gas, as traced by molecular hydrogen \citep[e.g.][]{Noterdaeme2008,Noterdaeme2011,Balashev2014} 
and/or 21-cm absorption \citep[e.g.][]{Srianand2012}. 
Important progress has been 
made towards understanding the properties of the gas (through e.g. deriving physical 
conditions \citealt{Srianand2005,Noterdaeme2007, Jorgenson2009} and physical extent \citealt{Balashev2011})
and the incidence of cold gas in DLAs has been related to other properties (such 
as the metallicity \citep{Petitjean2006} or the dust content \citep{Ledoux2003, Noterdaeme2008}).
However, due to the strong saturation of H\,{\sc i} Lyman series lines, it remains impossible to directly determine the H~{\sc i} column density associated with the individual cold gas components traced by H$_2$ absorption.
Even for metals, whose absorption lines are not saturated, it is very difficult to determine what fraction originates from the cold phase.
Difficulties arise as well with the 21-cm absorption that do not always exactly coincide 
with H$_2$ absorption \citep{Srianand2013} 
although it could be due to the different structures of the optical and radio 
emitting regions of the background quasars. 
Out of all the metals, chlorine shows a unique behavior in the presence of H$_2$.
Because the ionization potential of chlorine (12.97 eV) is less than that of atomic hydrogen, 
chlorine is easily ionized in the diffuse neutral medium. However, this species reacts exothermically with H$_2$ at a very high rate converting rapidly Cl$^+$ into HCl$^+$. The latter subsequently releases neutral chlorine through several channels \citep{Jura1974,Neufeld2009}. This process is so efficient that chlorine
is completely neutral in presence of a small amount of H$_2$. 
In our Galaxy the fact that chlorine abundance anti-correlates with the average number density along the line of sight \citep{Harris1984, Jenkins1986} has been interpreted as chlorine depletion. However, models predict as well as observations indicate, that gas with moderate dust content presents negligible depletion of chlorine \citep[e.g.][]{Neufeld2009, Savage1996, Jenkins2009}.
Observationally, a tight relation is indeed found between Cl~{\sc i} and H$_2$ in the local 
ISM \citep{Jura1974, Sonnentrucker2006, Moomey2012}. In this letter, we present the 
first study of this relation at high redshift and over a wide range of column 
densities. 

\section{Data sample and measurements}
Since the first detection by \citet{Levshakov1985}, about two dozen H$_2$ absorption 
systems have been detected at high redshifts in QSO spectra. 
The detection limit of the strongest Cl\,{\sc i} absorption line (1347\AA, f=0.0153 \citep{Schectman1993}) in high quality spectrum (${\rm S/N}\sim 50$, ${\rm R}\sim 50\,000$) corresponds to $N(\mbox{\ClI}) \sim 10^{12}$\,cm$^{-2}$. The solar abundance of chlorine is $10^{-6.5}$ that of hydrogen \citep{Asplund2009} and given previous measurements of N(\ClI)/2N(H$_2$) \citep[e.g.][]{Moomey2012} we conservatively limit our study to systems with $N({\rm H}_2) \gtrsim 10^{17}$~cm$^{-2}$.

\begin{figure}[t]
\begin{center}
        \includegraphics[clip=,width=0.95\hsize]{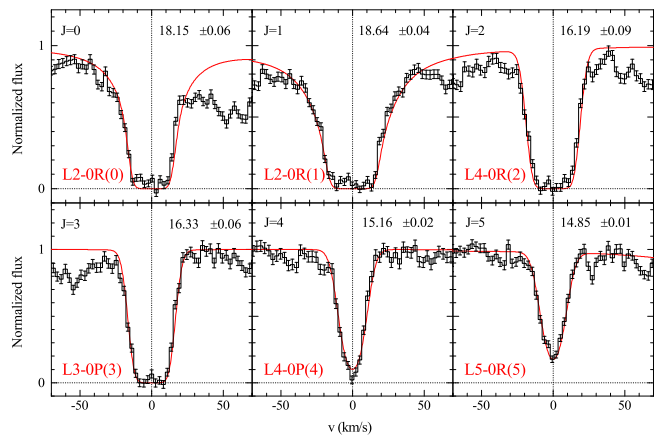}
        \caption{Voigt profile fits to the newly detected H$_2$ absorption lines from rotational levels J=0 to J=5 at $z=3.091485$ 
towards J\,2100$-$0641. The column densities are indicated (in log(cm$^{-2}$)) in the 
top right corner of each panel.}
        \label{J2100_H2}
\end{center}
\end{figure}

Redshifts, H\,{\sc i} and H$_2$ column densities, and metallicities were mainly taken from the literature 
and are based on VLT/UVES, Keck/HIRES or HST/STIS data.
We refitted H$_2$ absorption systems towards Q\,2123$-$0050 and Q\,J2340$-$0053 
to take into account the positions of the detected Cl~{\sc i} components.
We also detect a new H$_2$ absorption system in the $z=3.09$ DLA towards J\,2100$-$0641 in 
which \citet{Jorgenson2010} have reported the presence of neutral carbon. C\,{\sc i} is indeed known 
to be an excellent indicator of the presence of molecules \citep[e.g.][]{Srianand2005}. 
We used the MAKEE package (T. Burles) to reduce archival data from this quasar obtained in 2005, 2006 and 2007 under 
programs U17H (PI: Prochaska), G400H (PI: Ellison) and U149Hr (PI: Wolfe). We have found strong H$_2$ absorption 
lines from rotational levels up to $J=5$ (see Fig.~\ref{J2100_H2}) at $z=3.091485$ with a total column 
density of $\log N($H$_2) = 18.76 \pm 0.03$.

\begin{figure}[t]
\begin{center}
        \includegraphics[clip=,width=0.95\hsize]{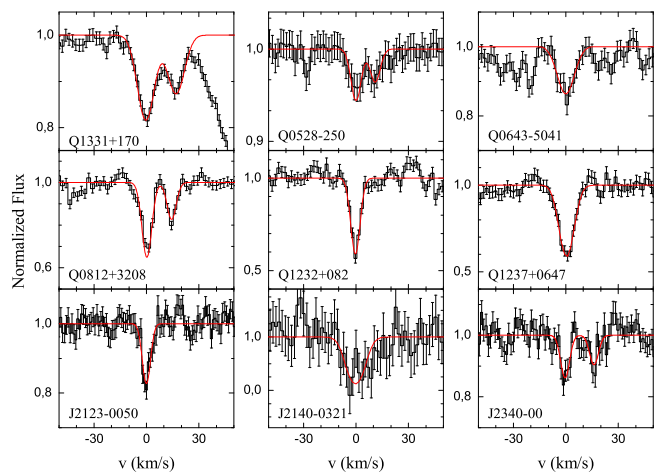}
  \caption{Voigt profile fits to \ClI\ $\lambda$1347 absorption lines associated with high-$z$ strong H$_2$ absorption systems. }
  \label{new}
\end{center}
\end{figure}

For all systems we retrieved data from the VLT/UVES or the Keck/HIRES archives.
We reduced the data and fitted the lines using profile fitting.
Neutral chlorine is detected in nine DLAs (Fig~\ref{new}). Four detections were already reported 
in the literature: Q\,1232+082 \citep{Balashev2011}, Q\,0812$-$3208 \citep{Prochaska2003}, 
Q\,1237$+$0647 \citep{Noterdaeme2010} and Q\,2140-0321 (Noterdaeme et al. submitted). The 
remaining five are new detections. 
We measured upper-limits of N(\ClI) for the remaining nine systems.
We used mainly the $1347$\AA\ \ClI\ line. 
Whenever possible, we also used \ClI\ lines at $1088$\AA, $1188$\AA, $1084$\AA, $1094$\AA\
and $1085$\AA, with oscillator strength from respectively \citet{Schectman1993}, \citet{Morton2003}, \citet{Morton2003}, \citet{Sonnentrucker2006} and \citet{Oliveira2006}.

Table~\ref{table_results} summarizes the results of \ClI\ measurements. We have kept all components with log~$N$(H$_2$)~$>$~17.
We did not use two known H$_2$ absorption systems towards Q\,0013$-$0029 and J\,091826.16$+$163609.0 since H$_2$ column densities in these systems are not well defined.

\begin{table*}
\centering
\addtolength{\tabcolsep}{-3pt}
\caption{Measurements of Cl\,{\sc i} in strong H$_2$ absorption systems at high redshift.        
\label{table_results}}
\begin{tabular}{l c c c c c c c c c c c }
\hline \hline
\multicolumn{1}{c}{Quasar} & {\large \strut} $z_{\rm em}$ & $z_{\rm DLA}$ & $\log N(\HI)$ & [X/H] & X & Ref. & $z_{\rm H2}$ & $\log N($H$_2)$ & $\log N(\ClI)$    & $b$ (km\,s$^{-1}$) &  \ClH\ \\
\hline
Q\, 0027$-$1836  & 2.56 & 2.40 & 21.75$\pm$0.10 &  -1.63$\pm$0.10 & Zn  & 1    & 2.40183   &    17.30$\pm$0.07        &  $<12.71$        & ---             &   $< 1.61$          \\
Q\, 0405$-$4418  & 3.02 & 2.59 & 21.75$\pm$0.10 &  -1.12$\pm$0.10 & Zn  & 2    & 2.59475   &    18.14$\pm$0.07        & $<12.71$         & ---             &   $< 0.77$           \\
Q\, 0528$-$2505  & 2.77 & 2.81 & 21.35$\pm$0.07 &  -0.91$\pm$0.07 & Zn  & 3    & 2.81098   &    18.11$\pm$0.02        &  11.92$\pm$0.08  & 4.1$\pm$1.5     &   0.01$\pm$0.08     \\ 
                &      &      &                &                 &     &      & 2.81112   &    17.85$\pm$0.02        &  11.73$\pm$0.11  & 4.2$\pm$2.0     &   0.08$\pm$0.11     \\
Q\, 0551$-$3637  & 2.32 & 1.96 & 20.70$\pm$0.08 &  -0.35$\pm$0.08 & Zn  & 4    & 1.96214   &    17.42$^{+0.45}_{-0.73}$  & $<12.40$         & ---             &   $< 1.54$         \\
Q\,J0643$-$5041  & 3.09 & 2.66 & 21.03$\pm$0.08 &  -0.91$\pm$0.09 & Zn  & 5    & 2.65860   &    18.54$\pm$0.01        &  12.51$\pm$0.05  & 5.8$\pm$1.4     &   0.17$\pm$0.05     \\
Q\,J0812$+$3208  & 2.7  & 2.63 & 21.35$\pm$0.10 &  -0.81$\pm$0.10 & Zn  & 6    & 2.62628   &    18.84$\pm$0.06        &  12.79$\pm$0.05  & 2.0$\pm$0.6     &   0.15$\pm$0.08     \\
                &      &      &                &                 &     &      & 2.62644   &    19.93$\pm$0.01        &  13.78$\pm$0.27  & 0.17$\pm$0.05   &   0.05$\pm$0.27     \\
Q\,J0816$+$1446  & 3.84 & 3.29 & 22.00$\pm$0.10 &  -1.10$\pm$0.10 & Zn  & 7    & 3.2874    &    18.62$\pm$0.18        & $<13.65$         & ---             &   $< 1.23$          \\
                &      &      &                &                 &     &      & 3.28667   &    17.60$\pm$0.27        & $<12.76$         & ---             &   $< 1.36$          \\
Q\, 1232$+$0815   & 2.57 & 2.34 & 20.90$\pm$0.08 &  -1.35$\pm$0.12 & S   & 8    & 2.33772   &    19.57$\pm$0.10        &  13.49$\pm$0.08  & 0.8$\pm$0.2     &   0.12$\pm$0.13     \\
Q\,J1237$+$0647  & 2.78 & 2.69 & 20.00$\pm$0.15 &  +0.34$\pm$0.12 & Zn  & 9    & 2.68959   &    19.20$\pm$0.13        &  13.01$\pm$0.02  & 4.5$\pm$0.4     &   0.01$\pm$0.13     \\
Q\, 1331$+$170   & 2.08 & 1.78 & 21.18$\pm$0.04 &  -1.22$\pm$0.10 & Zn  & 10,11& 1.77636   &    19.71$\pm$0.10        &  12.87$\pm$0.02  & 5.7$\pm$0.5     &   -0.64$\pm$0.10    \\
Q\,J1439$+$1118  & 2.58 & 2.42 & 20.10$\pm$0.10 &  +0.16$\pm$0.11 & Zn  & 12   & 2.4184    &    19.38$\pm$0.04        &  $<13.25$        & ---             &   $<0.07$           \\
Q\, 1441$+$2737  & 4.42 & 4.22 & 20.95$\pm$0.08 &  -0.63$\pm$0.10 & S   & 13   & 4.22401   &    18.05$\pm$0.05        & $<12.86$         & ---             &  $<1.01$            \\
                &      &      &                &                 &     &      & 4.22371   &    17.91$\pm$0.03        & $<12.66$         & ---             &  $<0.95$            \\
Q\, 1444$+$0126  & 2.21 & 2.09 & 20.25$\pm$0.07 &  -0.80$\pm$0.09 & Zn  & 14   & 2.08696   &    18.16$\pm$0.11        & $<12.42$         & ---             &   $< 0.46$          \\
Q\,J2100$-$0641  & 3.14 & 3.09 & 21.05$\pm$0.15 &  -0.73$\pm$0.15 & Si  & 15   & 3.09149   &    18.76$\pm$0.03        &  $<12.86$        & ---             &   $<0.3$            \\
Q\,J2123$-$0050  & 2.26 & 2.06 & 19.18$\pm$0.15 &  -0.19$\pm$0.10 & S   & 16   & 2.05933   &    18.09$\pm$0.02        &  12.27$\pm$0.06  & 2.6$\pm$0.5     &   0.38$\pm$0.06     \\
Q\,J2140$-$0321  & 2.48 & 2.34 & 22.40$\pm$0.10 &  -1.05$\pm$0.13 & P   & 17   & 2.33995   &    20.13$\pm$0.07        &  13.67$\pm$0.15  & 5-10            &   -0.26$\pm$0.18    \\     
Q\,J2340$-$0053 & 2.09 & 2.05 & 20.35$\pm$0.15 &  -0.92$\pm$0.03 & Zn  & 15   & 2.05456   &    18.07$\pm$0.06        &  12.25$\pm$0.15  & 0.8$\pm$0.6     &   0.38$\pm$0.16     \\
                &      &      &                &                 &     &      & 2.05473   &    18.25$\pm$0.05        &  12.13$\pm$0.18  & 1.0$\pm$1.0     &   0.08$\pm$0.19     \\
Q\, 2348$-$0108  & 3.01 & 2.43 & 20.50$\pm$0.10 &  -0.62$\pm$0.10 & S   & 18,19& 2.42688   &    18.12$\pm$0.37        &  $<13.86$        & ---             &   $< 1.94$          \\
                &      &      &                &                 &     &      & 2.42449   &    17.52$\pm$0.80        &  $<13.13$        & ---             &   $< 0.81$          \\
\hline
            \end{tabular}
          
\tablebib{(1) \citet{Noterdaeme2007}; (2) \citet{Ledoux2003}; (3) \citet{Noterdaeme2008}; (4) \citet{Ledoux2002}; 
(5) \citet{Albornoz2014}; (6) \citet{Jorgenson2009}; (7) \citet{Guimaraes2012}; (8) \citet{Balashev2011}; 
(9) \citet{Noterdaeme2010}; (10) \citet{Carswell2011}; (11) \citet{Balashev2010}; (12) \citet{Srianand2008};
(13) \citet{Ledoux2006}; (14) \citet{Ledoux2003}; (15) \citet{Jorgenson2010}; (16) \citet{Malec2010}; 
(17) Noterdaeme et al. submitted; (18) \citet{Petitjean2006}; (19) \citet{Noterdaeme2007a}. 
}
\addtolength{\tabcolsep}{3pt}
\end{table*}

\section{Results}

\begin{figure}
\centering
\includegraphics[bb=62 178 490 570,clip=,width=0.95\hsize]{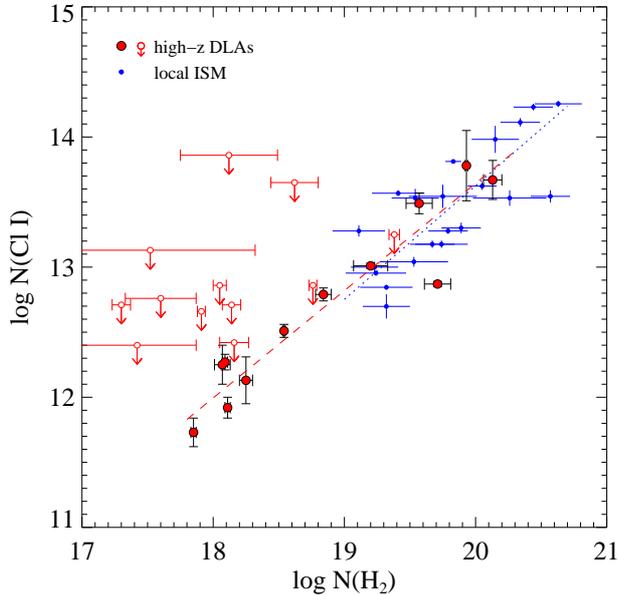} 
\caption{Column densities of Cl\,{\sc i} versus that of H$_2$. The red and blue points indicate the measurements at high redshift (this work) 
and in our Galaxy \citep{Moomey2012}, respectively. The straight (dashed, dotted) lines show the respective least-squares bisector fits to the data. \label{resn}}
\end{figure}

Fig.~\ref{resn} shows the Cl\,{\sc i} column density, $N$(\ClI), versus $N$(H$_2$) and 
compares our high-$z$ measurements to those obtained in the local ISM 
using the Copernicus satellite \citep{Moomey2012}. As can be seen our high-$z$ measurements extend the 
relation to H$_2$ column densities ten times smaller than those measured in the local ISM.
Cl\,{\sc i} and H$_2$ are found to be very well correlated ($r=0.95$) over the entire
$N$(H$_2$) range.
It is striking that measurements at high and low redshifts are indistinguishable in the overlapping regime ($\log N($H$_2) \sim $19-20.2).  
The correlation is seen over about three orders of magnitude in column density with a dispersion of 
about 0.2~dex only. A least-squares bisector linear fit provides a slope of 0.83 and 0.87 for the high-$z$ and 
$z=0$ data, respectively, with an almost equal normalization ($\log N(\ClI) \approx 13.7$ at $\log N($H$_2)=20$).
We note that the upper limits on $N$(\ClI) lie mostly at the low $N($H$_2)$ end and are least constraining
since they are compatible with the values expected from the above relation.
For this reason, we will not consider them further in the discussion but still include them in the 
figures for completeness. 
The slopes are less than one, meaning that the \ClI/H$_2$ ratio slightly 
decreases with increasing $N($H$_2)$.  This is unlikely to be due to conversion of Cl into H$_2$Cl$^+$ and/or HCl, since chlorine chemistry models \citep{Neufeld2009} as well as measurements (e.g. towards Sgr B2(S), \citealt{Lis2010}) show that in diffuse molecular clouds only $\sim1$\% of chlorine is in the molecular form. 
A $<1$ slope could in principle be due to dust depletion. However, there is no trend for increasing Cl depletion with increasing H$_2$ or \ClI\ column densities in Galactic clouds (see \citealt{Moomey2012}). 
In addition, for high redshift measurements, elemental abundance pattern \citep{Noterdaeme2008} as well as direct measurements \citep[e.g.][]{Noterdaeme2010} indicate $A_v < 0.2$ when modeling of chlorine chemistry \citep{Neufeld2009} shows that for such low extinction (A v < 1) all chlorine is in the gas phase. 	

A possibility is thus that the molecular fraction in the gas probed by \ClI\ is slightly increasing with 
increasing $N$(H$_2)$. It can be expected since H$_2$ self-shielding increase while \ClI\ is already completely in the neutral form. Finally, the similarity between our Galaxy and high-$z$ measurements at $\log N($H$_2)\ge 19$ might indicate that the chemical and physical conditions in the cold gas can be similar, otherwise fine tuning would be required between the different factors that impact on the $N$(\ClI) to $N$(H$_2$) ratio (e.g. number density, metallicity, dust content and UV flux).

Before continuing further, we note that in H$_2$-bearing gas, chlorine is found exclusively in the neutral form 
(i.e. $N($Cl$) = N(\ClI)$) \citep[e.g.][]{Jura1974}. 
Since we expect that all chlorine is in gas-phase\footnote{ We note that the presence of \ClII\ in the outer envelope of the H$_2$ cloud is not excluded, but it does not influence our derivation}, the abundance of chlorine, [Cl/H], in H$_2$-bearing gas can be expressed as  
\begin{equation}
{\rm [Cl/H]} = {\rm [\mbox{\ClI}/H_2]} + \log{f}, 
\label{Metalfl1}
\end{equation}
where 
\begin{equation}
{\rm [\mbox{\ClI}/H_2]} = \log\left( \frac{N({\mbox{\ClI})}}{2N({\rm H}_2)} \right)  - \log\left(\frac{{\rm Cl}}{{\rm H}}\right)_{\odot}
\label{ClH2}
\end{equation}
\noindent and $f=2N($H$_2)/(2N($H$_2)+N($H\,{\sc i}$))$ is the molecular fraction. 
Therefore the ratio \ClH\ gives a direct constraint on the chlorine-based metallicity of H$_2$-bearing gas provided 
the molecular fraction of H$_2$-bearing gas is known. 
Conversely, if a constraint can be put on the actual chlorine abundance, \ClH\ can provide an estimate of the amount of H\,{\sc i} present 
in H$_2$-bearing gas. 
If Cl is depleted onto dust grains then the mentioned estimates of metallicity and molecular fraction will have to be corrected from the Cl depletion factor.

\begin{figure}
\centering
\begin{tabular}{c}
\includegraphics[bb=62 178 490 570,clip=,width=0.8\hsize]{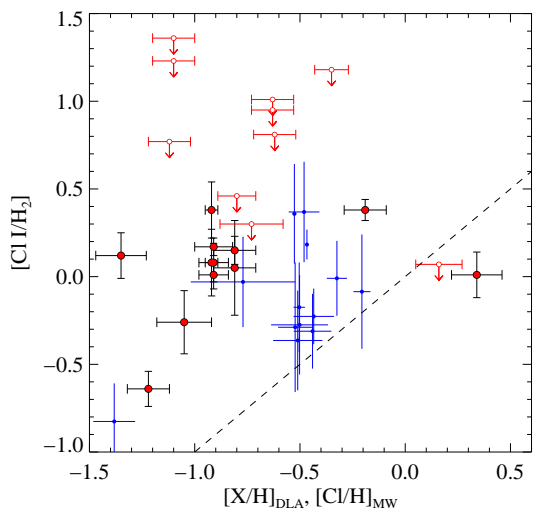} \\ 
\end{tabular}
\caption{\ClH\ as a function of the {\sl overall} metallicity for high-$z$ DLAs (red points) and the 
chlorine-based metallicity for Milky-Way clouds (blue points). The dashed line represents the one-to-one relation.
\label{res}}
\end{figure}

The Fig.~\ref{res} shows \ClH\ as a function of the
{\sl  overall} metallicity for DLAs (given in Table~\ref{table_results}) at high redshift or as a function of [Cl/H] for clouds in our Galaxy.
Since $f\le 1$, \ClH\ gives an upper limit on the metallicity in H$_2$-bearing gas,
which is found to be roughly equal or less than solar metallicity. For 13 out of 21 \ClI\ bearing clouds in our Galaxy 
associated \ClII\ was measured \citep{Moomey2012}. Therefore we have estimated the overall chlorine abundance [Cl/H] 
of these clouds as $(N(\mbox{\ClI}) + N(\mbox{\ClII}))/(N(\mbox{\HI}) + 2N(\mbox{H}_2))$ (shown by blue circles in Fig.~\ref{res}).
Unfortunately, for high redshift DLAs, not only \ClII\ is not detected but also  
DLAs contain several \HI\ clouds so that chlorine abundance of the very cloud of interest cannot be measured. Therefore we consider the overall metallicity (averaged over velocity components) measured using another non-depleted 
element (usually Zn or S, see Table~\ref{table_results}). In Fig.~\ref{res} it can be seen that the \ClH\ ratio is likely not correlated with the {\sl overall} metallicity of the DLA (Pearson correlation coefficient 0.3 at 0.3 significance level). 
For Milky-Way clouds it can be seen that [Cl/H] is typically one third solar, which can be interpreted 
as evidence for chlorine depletion \citep{Moomey2012}. 

The large difference between \ClH\ and [X/H]$_{\rm DLA}$ for the high redshift clouds can be explained either
by a molecular fraction $f<1$ in H$_2$-bearing clouds or by a higher metallicity in the 
H$_2$-bearing gas compared to the overall DLA metallicity or by both effects. If we assume that the 
metallicity in the H$_2$-bearing gas is equal to the overall DLA metallicity we find (using Eq.~\ref{Metalfl1}) 
that the molecular fraction in the H$_2$-bearing gas is typically an order of magnitude higher than 
the overall inferred DLA molecular fraction.
Interestingly, two systems sitting close to the one-to-one relation are those where CO molecules 
have been detected (Q\,1439+1118 and Q\,1237+0647). 
In such systems, the H$_2$ component is probably fully molecularized and its metallicity is close to the overall metallicity of the DLA.

\setlength{\tabcolsep}{0pt}
\setlength{\tabcolsep}{3pt}

\section{Conclusion}
\noindent

We have studied the neutral chlorine abundance in high redshift (z$\sim$2\,-\,4)
strong H$_2$ bearing DLAs with log~$N$(H$_2$)~$>$~17.3.  
These systems arise in the cold neutral medium of galaxies in the early Universe.
We have used 17 systems from the literature
and also present a new H$_2$ detection at $z=3.09145$ in the spectrum of J\,2100$-$0641. 
We have detected \ClI\ absorption lines in half of these systems, including 5 new detections. The derived upper limits for N(\ClI)
for the remaining systems are shown to be consistent with the behavior of the 
overall population.
Our measurements extend the \ClI-H$_2$ relation to lower column densities than 
measurements towards nearby stars. We show that there is a 5\,$\sigma$ correlation between the column densities of both 
species over the range 18.1~$<$~log~$N$(H$_2$)~$<$~20.1 with 
indistinguishable behavior between high and zero redshift systems. 
This suggests that at a given $N($H$_2$) the physical conditions are likely similar in our Galaxy and high-$z$ gas, in spite of possible differences in the dust depletion levels.
As we expect the Cl to be depleted less in the high-z absorbers studied here, we use the abundance of chlorine with 
respect to H$_2$ to constrain the molecular fraction and the metallicity in H$_2$-bearing gas. 
Our results suggest that the molecular fraction and/or the metallicity in the H$_2$ and \ClI\ bearing components could be 
much higher than the mean value measured over the whole DLA system.  This implies
that a large fraction of \HI\ is unrelated to the cold phase traced by H$_2$. 
Finally, our understanding of the formation of H$_2$ onto dust grains, self-shielding and lifetime 
of cold diffuse gas would certainly benefit from further observations of chlorine and molecular 
hydrogen in different environments and over a wide range of column densities. 

\vspace{2mm}{\footnotesize {\rm Acknowledgments.}
SB and VK thank RF President Program (grant MK-4861.2013.2) and ``Leading Scientific Schools of Russian Federation'' (grant NSh-294.2014.2). 
RS and PPJ gratefully acknowledge support from the Indo-French Centre for the
Promotion of Advanced Research (Centre Franco-Indien pour la
Promotion de la Recherche Avanc\'ee) under contract No. 4304-2.
\bibliographystyle{aa}
\bibliography{H2}

\end{document}